\documentclass[preprint,nonacm]{acmart}
\usepackage{balance}
\usepackage[utf8]{inputenc}
\usepackage{amssymb}
\usepackage{lineno}
\usepackage[english]{babel}
\usepackage{palatino,url}
\usepackage{amsmath}
\usepackage{tcolorbox}
\usepackage{xcolor}
\usepackage{listings}
\usepackage{graphicx}
\usepackage{subfig}
\usepackage[rightcaption]{sidecap}
\usepackage{textcomp}
\usepackage{array} 
\usepackage{booktabs} 
\usepackage{multirow}

\definecolor{codegreen}{rgb}{0,0.6,0}
\definecolor{codegray}{rgb}{0.5,0.5,0.5}
\definecolor{codepurple}{rgb}{0.58,0,0.82}
\definecolor{backcolour}{rgb}{0.95,0.95,0.92}
\definecolor{lightgray}{gray}{0.9}   
  
\definecolor{yellow}{RGB}{255,255,153}
\definecolor{grey}{RGB}{224,224,224}

\newboolean{showcomments}
\setboolean{showcomments}{true}
\ifthenelse{\boolean{showcomments}}
 { \newcommand{\mynote}[2]{
      \fbox{\bfseries\sffamily\scriptsize#1}
        {\small$\blacktriangleright$\textsf{\emph{#2}}$\blacktriangleleft$}}}
        { \newcommand{\mynote}[2]{}}

\definecolor{DarkOrange}{rgb}{0.8,0.3,0.0} 
\definecolor{DarkCyan}{rgb}{0.0, 0.55, 0.55}
\definecolor{Gray}{gray}{0.9}

\newcommand{\find}[1]{\begin{tcolorbox}[leftrule=1mm,rightrule=1mm,toprule=0mm,bottomrule=0mm,left=2pt,right=2pt,top=0pt,bottom=0pt]
#1
\end{tcolorbox}
}

\AtBeginDocument{%
  \providecommand\BibTeX{{%
    \normalfont B\kern-0.5em{\scshape i\kern-0.25em b}\kern-0.8em\TeX}}}

\copyrightyear{}
\acmYear{}
\setcopyright{none}
\acmConference{}
\acmBooktitle{}
\acmPrice{}
\acmDOI{}
\acmISBN{}

\begin{document}

\title{Data-driven  Simulation and Optimization for Covid-19 Exit Strategies}


\author{Salah Ghamizi, Renaud Rwemalika, Maxime Cordy, Lisa Veiber, Tegawendé F. Bissyandé, Mike Papadakis, Jacques Klein and Yves Le Traon}

\affiliation{
  \institution{\\SnT, University of Luxembourg}
}
\email{email: {firstname.lastname}@uni.lu}
\renewcommand{\shortauthors}{Ghamizi, Rwemalika et al.}

\begin{abstract}
The rapid spread of the Coronavirus SARS-2 is a major challenge that led almost all governments worldwide to take drastic measures to respond to the tragedy. 
Chief among those measures is the massive lockdown of entire countries and cities, which beyond its global economic impact has created some deep social and psychological tensions within populations. While the adopted mitigation measures (including the lockdown) have generally proven useful, policymakers are now facing a critical question: {\em how and when to lift the mitigation measures?} A carefully-planned exit strategy is indeed necessary to recover from the pandemic without risking a new outbreak. Classically, exit strategies rely on mathematical modeling to predict the effect of public health interventions. Such models are unfortunately known to be sensitive to some key parameters, which are usually set based on rules-of-thumb.  

In this paper, we propose to augment epidemiological forecasting with actual data-driven models that will learn to fine-tune predictions for different contexts (e.g., per country). We have therefore built a pandemic simulation and forecasting toolkit that combines a deep learning estimation of the epidemiological parameters of the disease in order to predict the cases and deaths, and a genetic algorithm component searching for optimal trade-offs/policies between constraints and objectives set by decision-makers.

Replaying pandemic evolution in various countries, we experimentally show that our approach yields predictions with much lower error rates than pure epidemiological models in 75\% of the cases and achieves a 95\% R² score when the learning is transferred and tested on unseen countries. 
When used for forecasting, this approach provides actionable insights into the impact of individual measures and strategies. 
\end{abstract}



\maketitle

\pagestyle{plain}
\section{Introduction}


Since the outbreak of the COVID-19 pandemic, the world has been facing a human tragedy with overwhelmed healthcare systems and fears of economic collapses. In the absence of vaccines to immunize the population rapidly at scale, governments have implemented various non-pharmaceutical public health interventions such as social distancing and lockdowns. Considering that the World Health Organisation (WHO) is foreseeing first clinical trials of vaccine for the end of the year 2020~\cite{WHO2020a}, decision-makers must carefully plan their exit strategies: measures that were put in place to contain the coronavirus spread must be methodically lifted to avoid the risk of precipitating new outbreaks. 


 
In this context, mathematical modelling offers public health planners with frameworks to make predictions about the spread of emerging diseases and assess the impact of possible mitigation strategies. This is particularly important when dealing with infectious diseases, such as COVID-19, where mass interventions (\emph{e.g.,} screening, social distancing, and vaccination) can lead to effects at a population level, including herd immunity, changes in the infection rate or even changes in the pathogen ecology as a consequence of selective pressure.

There are two main types of models: static cohort models and transmission dynamic models~\cite{Jit2011}. Static models, typically relying on decision trees and Markov processes, assume a force of infection that is independent of the proportion of the population that is infected and therefore is of little use in response to highly infectious diseases like COVID-19. Transmission dynamic models, on the other hand, see a force of infection varying depending on the proportion of the population which is infected. Compared with static cohort models, transmission dynamic models are usually more complex to parameterize requiring epidemiological information on the infectious disease and demographic and economic information about the affected population.

Different techniques exist to implement dynamic approaches. Agent-Based Models (ABM) are simulations composed of agents that interact with each other and their environment. Because each agent can make its own rules, this type of approach can capture aggregate phenomena derived from the behavior of single agents. These models offer a great explainability of the root causes leading to the propagation of a disease but are computationally intensive to run and thus, hardly applicable to large populations. 
Indeed, the behaviour and the interaction of each type of agent needs to be fully defined in order for the model to be useful. These rules are case-specific and are not transferable from one population to another.

The most common approach to model the spread of infectious disease is the Susceptible-Infected-Removed
(SIR) model and its extension \emph{i.e} SEIR (Susceptible, Exposed, Infectious, and Recovered). This is a state-based model, every state expresses the degree of exposure of a population to the disease.
It is also equation-based where each equation defines the rate to go from one state to the other. The SEIR model thus separates the population into four groups and simulates the evolution over time of each one of the subpopulations. The transition rates are defined by the time scale to which an individual can transmit the disease, the time to recovery, and the number of newly infected people due to an infected individual. The most varying parameter is called effective reproduction number ($R_t$), and expresses the number of people that can be contaminated by an infectious individual over a period of time. 

These methods are dependent on the validity of the input parameters like transition rates. While SEIR is a very powerful model, it presents a major limitation, it requires hyper-parameters that are hard to observe such as the infection rate of an individual.
In practice, SEIR parameters are manually set to fit with the local observations to the considered population (e.g. country), and are not learnt from larger-scale observations. To circumvent the limitation of such epidemiological models, researchers recently started to take advantage of the advances made in Machine Learning (ML) in order to create models based on available large datasets \cite{Vollmer2020,Soures2020}. We name this family of approaches ``ML-based'' and our own work falls in it.

Our first contribution is to devise a novel approach, \emph{DN-SEIR}, that alleviates manual tuning of the SEIR model, by relying on large and trustable public datasets (large scale observations) and machine learning (to learn the parameters' values for a given population). Our approach combines SEIR with a machine learning predictor, based on a deep learning model, to estimate the effective reproduction number ($R_t$), over time. The machine learning predictor relies on demography and mobility features to predict an effective reproduction number. For each time increment, $R_t$ is updated and used for the next day computation. We evaluated our approach on twelve countries from all continents and showed that our approach that mixes demographics, mobility, and epidemiological data provides better forecasts for 9 out of 12 of the studied countries than a purely epidemiological modelling.

Our second contribution is to exploit this online prediction of effective reproduction number in a simulation tool\footnote{Open sourced and available for reproduction on https://github.com/covid19-kdd/kdd20covid19/} for policymakers which was recently advertised to the public \footnote{Online tool and press release anonymized for double-blind review}. Policymakers have to decide when to relax certain parts of society (workplaces, travels, schools... ) and to what extent it may create a new epidemic wave that would flood the hospitals with critical cases. The simulator enables one to make such a strategic exit plan for a certain country and predict its impact in terms of hospitalization, infected people and deaths. It is also designed to explore and optimize various exit strategies and constraints. We evaluate 3 common hand-crafted exit strategies and show that Multi-objective genetic algorithms can find atypical strategies on the pareto-front that minimize both the death numbers and the economic impact. 
\vspace{-1mm}
\section{Related work}
\vspace{-1mm}
Machine Learning approaches have been widely used to model and forecast former epidemics, especially to handle the large amount of data it involves and the increasing complexity of the epidemiological models underneath. Popular approaches remain regression trees and forests~\cite{dengue} and neural networks~\cite{h1n1}. While there is a plethora of reports that tackle the COVID-19 forecasting, the peer-reviewed literature about ML and COVID-19 is rather scarce. Most approaches in public repositories tackle ML regression in combination with the SIR epidemiological model \cite{ardabili2020covid} or its SEIR extension~\cite{ pandey2020seir, yang2020modified, 2020sirnet}. 

Recent research about non-pharmaceutical interventions focused on mobility data to combine epidemiological models and learning algorithms. 
In~\cite{vollmer2020report}, Vollmer et al. integrate mobility in a stochastic model. They focused on Italy and suggest that COVID-19 transmission rate and mobility metrics are closely related.
as well as mobility should be closely monitored in the next weeks and months.. 

Our approach relies on a similar intuition and combines learning from multiple countries with country-specific features. The transfer learning in our approach is able to separate the contribution of different measures like commute to work, retail and recreation activities. We also show that using the output of our ML approach as a search fitness function leads to optimal exit strategies. 

\section{Approach}
\label{section:methods}

\begin{figure*}
\centering
\includegraphics[width=0.8\linewidth]{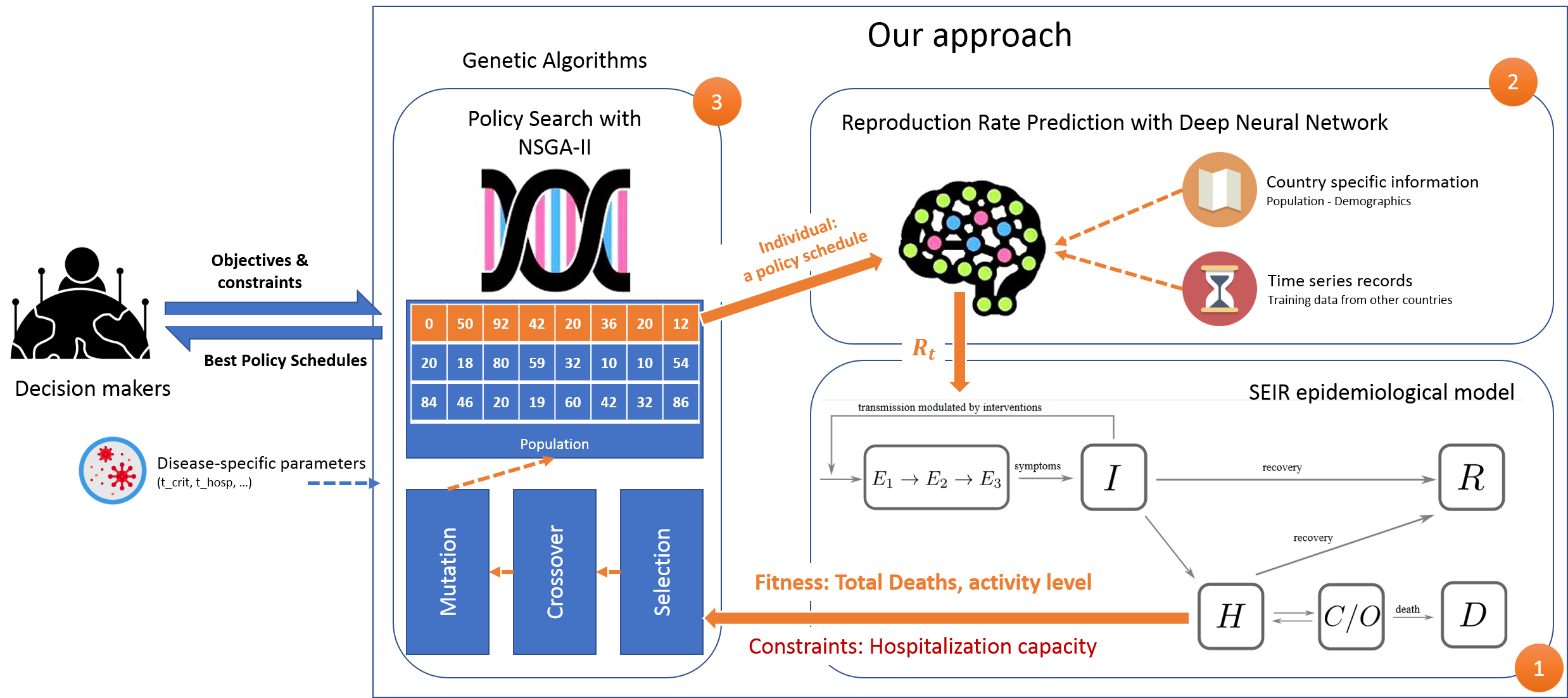}
\caption{Our approach relies on a feedback loop where the Genetic Algorithm searches for optimal exit strategies using a fitness function computed from the epidemiological model outputs. The epidemiological model's parameters are  learned with a Machine Learning algorithm that uses population mobility behaviours and demographics as input features.}
\label{fig:approach}
\end{figure*}

Our end goal is to provide policymakers with a tool to easily generate exit strategies and evaluate their impact. In particular, an exit strategy can be modeled like as a schedule of measures (\emph{policy schedule}) that will impact the way the disease will spread. We restrict the policy schedule to mobility levels.

As illustrated in Figure~\ref{fig:approach}, we propose to combine a genetic algorithm (to search for policy schedules), a deep learning model (to predict the evolution of the effective reproduction number induced by a given policy schedule) and an epidemiological model (to forecast, based on the computed effective reproduction numbers, the effect of the scheduled policies on public health over time, e.g. deaths and hospitalization occupancy). Our three components work within a feedback loop. At each iteration, the genetic algorithm builds a population of policy schedules, which the deep learning and epidemiological models allow to evaluate. In turn, this feedback is used to generate better schedules, optimizing health-related objectives (e.g. minimize total deaths) while satisfying hard constraints (e.g. never exceed hospitalization capacity).

\subsection{Estimating Impacts on Public Health with Epidemiological Models}

Epidemiological models predict the state of a population struck by a pandemic over time, based on state transition parameters and the evolution of the effective reproductive number, $R_t$, of the disease.

We use an extension of SEIR model, i.e. the SEI-HCRD compartmental model -- Susceptible ($S$)  $\rightarrow$ Exposed ($E$) $\rightarrow$ Infectious ($I$) $\rightarrow$ Removed (Hospitalized ($H$), Critical ($C$), Recovered ($Rec$), Dead ($D$)). Such model can be defined by the following system of ordinary differential equations:

\begin{align}
    \frac{dS}{dt} &= - \frac{R_t}{t_{inf}} \cdot I \cdot S \\        
    \frac{dE}{dt} &=  \frac{R_t}{t_{inf}} \cdot I \cdot S - \frac{1}{t_{inc}} \cdot E \\
    \frac{dI}{dt} &= \frac{1}{t_{inc}}  \cdot E - \frac{1}{t_{inf}} \cdot I \\
        \frac{dH}{dt} &= \frac{1-m}{t_{inf}}  \cdot I + \frac{1-f}{t_{crit}} \cdot C - \frac{1}{t_{hosp}}  \cdot H 
    \\ 
    \frac{dC}{dt} &= \frac{c}{t_{hosp}} \cdot H - \frac{1}{t_{crit}} \cdot C    \\ 
    \frac{dRec}{dt} &= \frac{m}{t_{inf}}  \cdot I + \frac{1-c}{t_{hosp}} \cdot H    \\ 
    \frac{dD}{dt} &= \frac{f}{t_{crit}}  \cdot C
\end{align}
such that $S + E + I + H + C + Rec + D$ is equal to the total population and $R_t$ denotes the effective reproduction number over time. 

The SEI-HCRD involves several parameters. $t_{suffix}$ is the transition time estimated to transit from one population to the other. $t_{inc}$ is the average incubation period, $t_{inf}$ is the average infectious period, $t_{hosp}$ is the average hospitalization time in normal state (i.e. until any patient recovers or enters a critical state). $t_{crit}$ is the average hospitalization time in a critical state (i.e. until death or recovery). The parameters $m$, $c$, $f$ determine the severity of the infection: $m$ is the percentage of infected individuals with non-severe symptoms (i.e. they are asymptomatic or have mild symptoms) and which, therefore, are not hospitalized. $c$ is the percentage of hospitalized persons who will eventually enter a critical state. Finally, $f$ denotes the percentage of persons in the critical state who will pass away.

\begin{table}[h]
    \begin{tabular}{|cr||cr|}
        \hline
        \textbf{Transition time} & \textbf{Value} & \textbf{Transition ratio} & \textbf{Value}\\
        \hline
        $t_{inc}$ & 5.6 days & $m$ & 80\% \\
        $t_{inf}$ &  2.9 days & $c$ & 10\% \\
        $t_{hosp}$ & 4 days & $f$ & 30\% \\
        $t_{crit}$ & 14 days & & \\
        \hline
    \end{tabular}
    \caption{Parameter values used in the SEI-HCRD model.}
    \label{table:seir_parameters}
\end{table}

To set these parameters, we lean on Liu et al.'s study \cite{Liu2020} and assign them with the constant values reported in Table~\ref{table:seir_parameters}. Then, given the time series of effective reproduction numbers over time,$\{R_t\}$, the SEI-HCRD model computes the resulting impacts on the population, including the number of deaths. Instead of manually assigning fitted values to $R_t$, we propose to predict them from scheduled exit strategies using deep learning models.

\subsection{Predicting the Effective Reproduction Number over time with Deep Learning}



\subsubsection{Feature engineering}
\label{section:features}

We start from Google's Mobility Reports\footnote{https://www.google.com/covid19/mobility/}, which track the mobility trends over time, for different categories of places in 97 different countries. Each feature corresponds to a category of places and its value captures the daily traffic of such places. More precisely, the value of the feature is the difference between the daily traffic and the traffic baseline (i.e. the median traffic for the same weekday during the 5-weeks ranging from January 3 to February 6, 2020). The reports include 6 categories:  Grocery \& pharmacy, park, transit stations, retail \& recreation, residential and workplace.


\begin{figure*}
\centering
\subfloat[Luxembourg]{\includegraphics[width=0.31\linewidth]{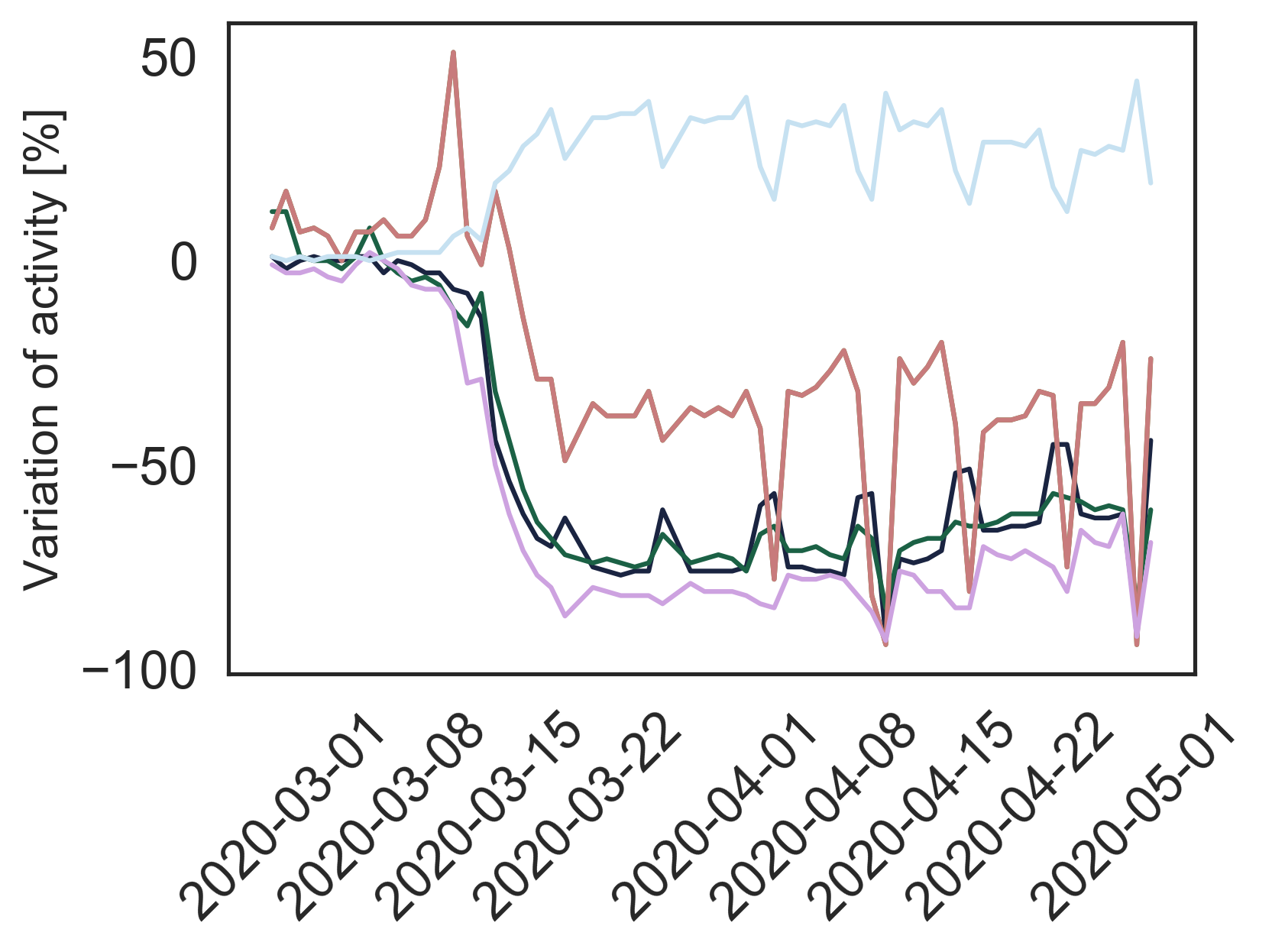} \label{fig:mobility_luxembourg}}
\subfloat[Italy]{\includegraphics[width=0.31\linewidth]{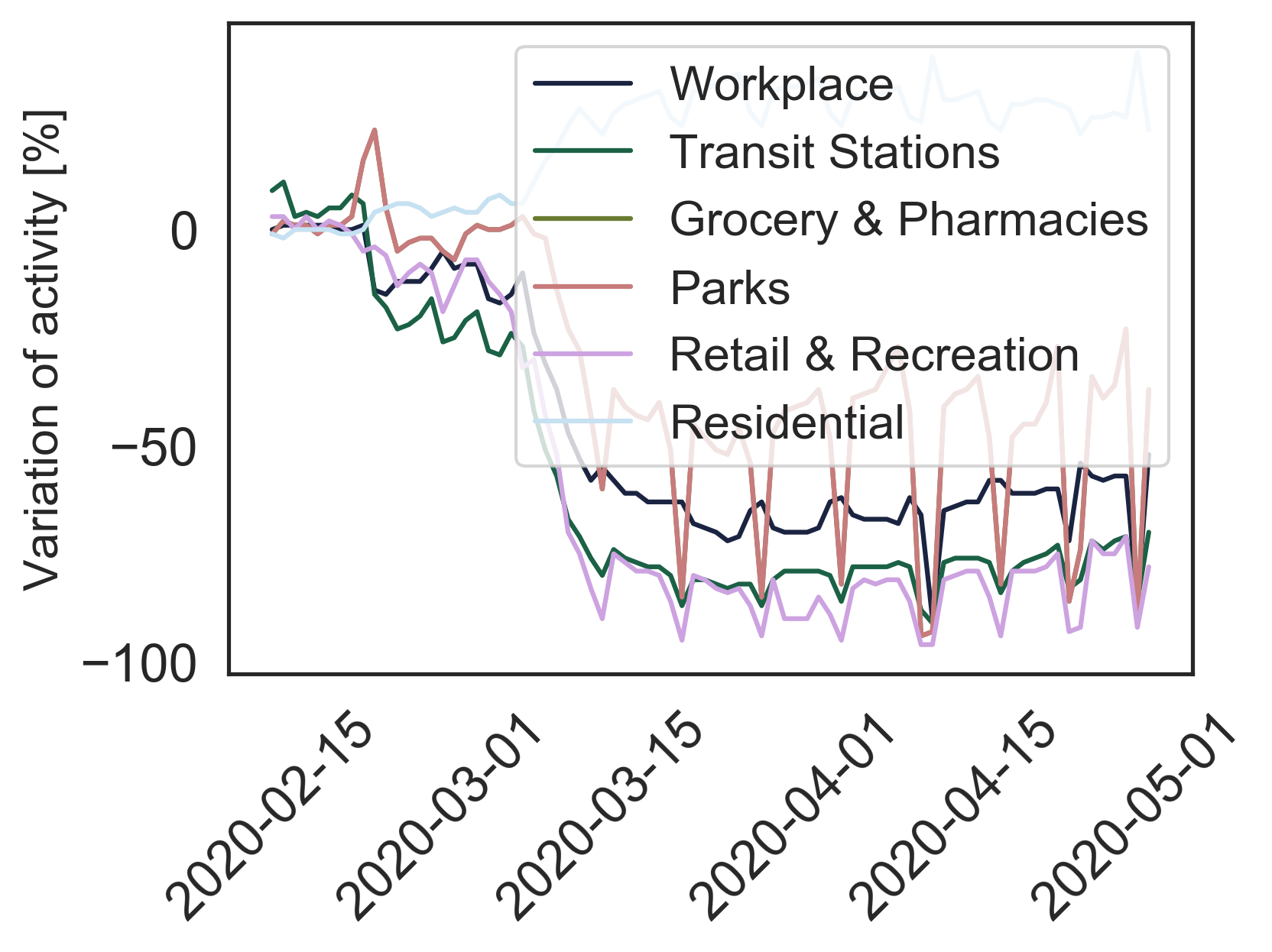} \label{fig:mobitlity_italy}}
\subfloat[Japan]{\includegraphics[width=0.31\linewidth]{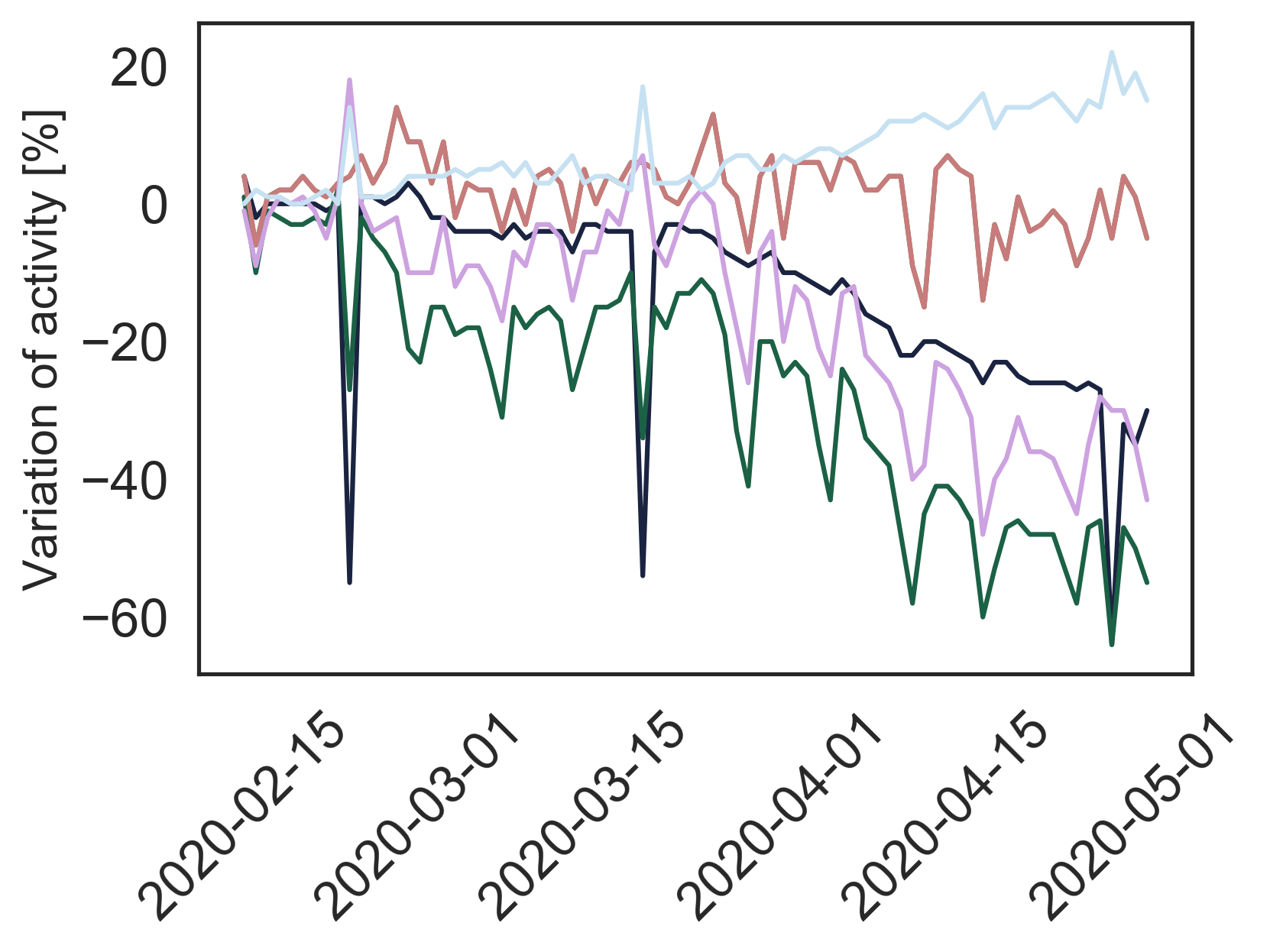} \label{fig:mobility_japan}}
\caption{Evolution of the mobility indicators for Luxembourg, Italy, and Japan. A value of 0 means that the activity is at the same level as before the confinement, a value of -100\% is a total stop of the activity and a positive values shows an increase of the activity compared to the reference value.}
\label{fig:mobility}
\end{figure*}

Of course, the values of these features are country-specific and are largely impacted by the mitigation strategy of each country. For example, Figure~\ref{fig:mobility} shows the evolution of all features for Luxembourg, Italy and Japan. We observe that Italy and Luxembourg have drastically reduced their activities whereas Japan does not exhibit a significant reduction, for the case of schools and international travels. 

Next, we clean the collected data. When some values (for a given feature) are missing, we fill the gap by interpolating between the closest days with available information. For each category of places, we smooth the corresponding feature over the 5, 10, 15, and 30 past days, resulting in four new features. 

We complete our dataset with demographic features and with the corresponding day of the week to take into account the weekly fluctuations of the data. 

Overall, our feature engineering process yields 4,625 inputs of 32 features each, which are recapitulated in Table \ref{table:dnn_datasets}.

\begin{table}[h]
\centering
\begin{tabular}{|l|l|r|r|}
\hline
\textbf{Feature category} & \textbf{Feature} & \textbf{Values}\\
\hline
Mobility & Grocery \& Pharmacy &  [-100,50]\\
& Parks & [-100,50]\\
& Residential & [-100,50]\\
& Retail \& Recreation & [-100,50]\\
& Transit stations & [-100,50]\\
& Workplace & [-100,50]\\
\hline
Demographics & GDP  & ${\rm I\!N}$\\
& Population  & ${\rm I\!N}$\\
& Density  & ${\rm I\!R}$\\
& Area &  ${\rm I\!N}$\\
& Proportion of population & \\
& under 15yrs  & [0,1]\\
& Proportion of population & \\
& over 64yrs & [0,1]\\
& Day of week & [0-6]\\
& Region (Continent) & [0-10]\\
\hline
\end{tabular}
\caption{Features of the Feed Forward Neural Network. Mobility features are augmented by smoothing over 5, 10, 15 and 30 days, hence each mobility feature corresponds to 4 inputs.}
\label{table:dnn_datasets}
\end{table}


\subsubsection{Training \& validation}


The model can be seen as a supervised predictor, taking as input the mobility and demographic features to predict an effective reproduction number, $R_t$, for each time index $t$. Thus, in order to train it, we need to label the dataset with $R_t$ value for each day of the training period. 

Since the real effective reproduction numbers are not known, we estimate them by fitting the SEI-HCRD model to the real-world number of cases and deaths. Since the $R_t$ are time-dependent, we represent them as a decaying function and seek the best parameters' value for this function (which yield the fittest SEI-HCRD model). We opted for the Hill decay function because it showed good results in the literature \cite{malaria}. Thus, $R_t$ is given by $\frac{1}{(1 + (\frac{t}{L})^k)} \cdot R_0$ where $R_0$, $k$ and $L$ are parameters. We use the L-BFGS optimization method to find the values of these parameters that minimize the Mean Square Error of the number of cases and deaths predicted by the SEI-HCRD model. Once these parameter values are found, they can be injected back in the Hill decay function to generate a time series for the past values of $R_t$. 



The analysis of feature correlations shows that when working on a country-by-country basis, all the mobility features are highly correlated (over 0.75) for some countries as all the activity reduction and closure were enacted in most sectors at the same time. Co-linear features offering little to no information gain to the learner, we train the model on all countries to reduce any correlation between the different features. 
Thus, the model is trained using all the countries at once and the train/test split is done randomly.

Once a training set with expected values ($R_t$ values computed from the estimated Hill decay function) is established, a supervised model can be trained to predict the future values of $R_t$ using the mobility and demographic data as features.

To do so, we rely on a  Feed-Forward Neural Network (FFNN). The architecture of the FFNN and its hyper-parameters are optimized using a grid search to minimize the mean square error with cross-validation. The search leads to an architecture with 2 fully-connected hidden layers with, respectively, 1000 and 50 neurons.

In addition to the FFNN, we also evaluate two other estimators. Since the problem takes the form of a time series, we investigate the performances of a Long Short-Term Memory (LSTM) network which allows sequence to sequence transformation, hence, learning from the features of the past days to predict the next days. We use a 15 days window to predict the values for the next 7 days. Note that in this case, at each step, we use the computed values of $R_t$ from the past iterations as input for the model in addition to the rest of the features. Finally, we investigate one last approach, Gradient Boosting using 500 estimators. For each of the approaches, we evaluate the performances using two classical metric, \emph{i.e} the coefficient of determination, ($R^2$ score) and the Root Mean Square Error (RMSE)\footnote{The closer $R^2$ score to 1 the better and the closer RMSE to 0 the better}. The train:test splits were performed with (1) random split between train and test data (2) split based on the region the country is located in (i.e testing on unseen countries) (3) split based on the time, all values before a certain date are considered for the training set and the ones after are used to build the test set. In all instances, we keep a ratio train:test around 80:20.

The FFNN provided a $R^2$ score of 0.95 with a random split and 0.97 with a region split, while the gradient boosting could only achieve a $R^2$ score of 0.83. The LSTM offered slightly better performances with a $R^2$ score of 0.95 when splitting on a region base to over 0.99 when splitting on a time base. We use the FFNN model in our following experiments. We refer to the combination of FFNN and SEI-HCRD as \textbf{DN-SEIR}.






\subsubsection{Interpretability}

Although Machine Learning algorithms provide increased accuracy in a wide variety of domains, their black-box nature makes them inherently non-interpretable. Indeed, in our case, a multi-layer neural network solely contains information in the form of numerical weights and connections. 
The model reasoning from input to output remains opaque. 
Nevertheless, interpretability can be reached as a post-hoc analysis through an independent interpretability framework. We choose Shapley Additive exPlanations, \textit{SHAP} \cite{SHAP} as it provides intuitive visualization-based explanations which can be incorporated in the simulator.  
The \textit{SHAP} framework is based on game theory and Shapley value. In game theory, Shapley values indicate how to fairly distribute a ‘payout’ among players. A model can be thought of as a game where each feature value for each instance is a player, while a prediction or model output is a payout. 
In practice, the Shapley value of a feature value is the contribution of this value for this particular exemplar compared to the average prediction for the specific dataset. 
\textit{SHAP} provides several advantages. First, this framework has different types of explainers to optimally provide explanations to different models, whether tree-based or kernel-based. Moreover, \textit{SHAP} exhibits an Efficiency property through the Shapley values. Indeed this component guarantees that the difference between prediction and average prediction is fairly distributed among the values of the features of this particular prediction. 


\subsection{Optimization of Policy Schedules with Genetic Algorithms}


Our search method uses NSGA-II \cite{nsga2}, an established Genetic Algorithm (GA) for multi-objective optimization that uses non dominated sorting to find pareto-optimal solutions. 

\paragraph{Solution space:} Any solution generated by NSGA-II is a policy schedule.  A schedule consists of a list of vectors, each of which is associated with a mobility feature and encodes the value of the feature for each time index $t$. A value ranges from 0 (no restriction) to 100 (full lockdown). The indices $t$ go from April 30 to September 30, with steps of 2 weeks. 

\paragraph{Objectives:} We use 2 fitness functions which represent the compromising health and societal impacts of policy scheduled. We quality these impacts with the total number of total deaths between April 30 and December 30 and the mean of the mobility feature values over the same period. The first objective must be minimized while the second is maximized.

\paragraph{Constraints:} For a policy schedule to be an acceptable solution, we require that the number of critical cases never exceeds the hospitalization capacity (ICU) of the country of interest. This is an important requirement for policymakers, as critical cases which may not be correctly hospitalized likely result in additional deaths.

\paragraph{Selector:} Current Pareto-front solutions are selected in priority. If there are more than the population size, they are filtered based on a crowding distance (here, we use Manhattan distance in the objective space). Otherwise, we fill the population with non-optimal solutions, selected using a binary tournament selection: Pareto-dominant solutions are retained in priority; in case of non-dominance, crowding distance to the Pareto front is used for tie-breaking.

We rely on Pymoo \footnote{www.pymoo.org, the most starred python GA library on Github} to implement our NSGA-II search and use the library's default values for the remaining parameters like mutation rate or crossover. 

\section{Research questions}
\label{section:rqs}

Our end goal is to provide decision-makers with a tool allowing to easily generate exit strategies in order to evaluate their impact. To achieve this, we use a deep neural network as a proxy to evaluate the hyper-parameters of a SEI-HCRD model based on mobility and demographic data. However, to be useful, the neural network needs to be able to capture enough information in the mobility and demographic data alone to make accurate predictions. Hence, we formulate the first research question as follow:





\find{\textbf{RQ1:} Can we predict the effective reproduction number based on mobility and demographic data?}

Ultimately the proposed approach is intended to allow policymakers to evaluate and select exit strategies by analysing their impact on multiple aspects such as the number of death, possible overflow of healthcare capacities or the perturbation of economic activities. To evaluate the capacity of our approach to model such strategies, we evaluate it by predicting the impact of various popular scenarios. Thus, we ask the following question:


\find{\textbf{RQ2:} How does our approach react under different exit strategies?}

We conclude our investigation by a comparison of the impact of the popular scenarios with the impact of the one proposed by the search algorithm. The algorithm minimizes the number of deaths and the socio-economical impacts generated by a diminution of activities (mobility) while avoiding the over-saturation of healthcare capacities. To evaluate the results, we compared them to the ``naive'' scenarios formulated in the previous research question and thus ask:

\find{\textbf{RQ3:} How do the exit strategies proposed by the search algorithm perform against popular ones?}
\section{Results}
\label{section:results}

\subsection{Predicting the Effective Reproduction Number}

\paragraph{Comparison of a fitted SEIR model and our DN-SEIR model}
We estimate the confidence interval by evaluating the mean and standard deviation of a Bayesian Ridge Regressor on each element of the test set (using the same training set as the FFNN). We use a grid search over 3 values for each of its 2 hyper-parameters $\alpha_{init}$ (0.1, 1, 1.9) and $\lambda_{init}$ (1, 0.1, 0.01). 


We compare in Table~\ref{table:shortterm_fitting_cases} the predicted cases of the time-dependant SEIR (Fitted on past cases/deaths) and the DN-SEIR approach. 
The DN-SEIR approach has a lower error to the ground truth values of cases for 9 over 12 countries in comparison with a time regression approach. Besides, the ground truth falls within the confidence interval of the DN-SEIR model for 10 of the 12 countries evaluated. It is worth noting that while 5 of the 12 countries lie under 5\% error, 9 over 12 countries do no exceed 15\% error.

\begin{figure}
\centering
\includegraphics[trim=0 5mm 0 5mm, clip, width=0.5\linewidth]{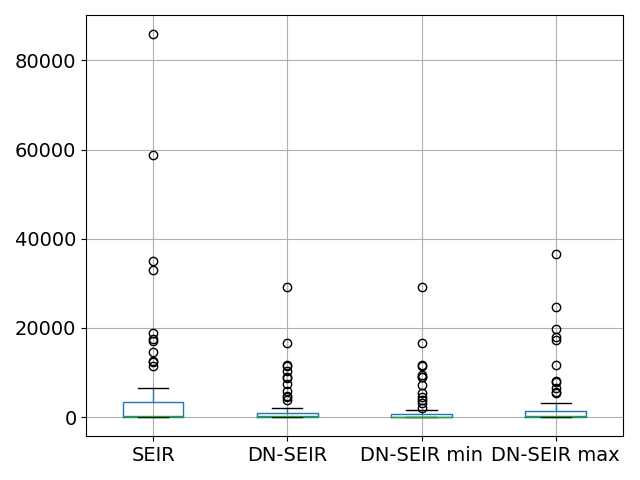}
\caption{RMSE between true cases and predicted cases by our DN-SEIR model and a fitted SEIR model. }
\label{fig:rmse_all}
\end{figure}

In Figure \ref{fig:rmse_all}, we evaluate the RMSE between the predicted cases of the SEIR model and the DN-SEIR approach (with its optimistic and pessimistic boundaries) across all the countries of the dataset. The simulation spans on 7 days (instead of 12 days for the results presented in table~\ref{table:shortterm_fitting_cases}) and we compare the number of cases on April 29th, the last day of available mobility data for all countries. We use the  Wilcoxon signed-rank test to compare if two distributions are equals. The results show that for all three prediction DN-SEIR, DN-SEIR max and DN-SEIR min we can reject the hypothesis (\emph{p-values} << 0.05) that they are equal to the SEIR prediction. We then perform a Vargha and Delaney's $A_{12}$ test to analyze the effect size. We see that our approach generates a lower RMSE with a small effect size. These results indicate that even in very short term (7 days), relying only on the mobility data yields better results than applying a regression over the past values.


\begin{table*}[ht]
\centering
\begin{tabular}{|l|r|r|r|r|r|}
\hline
\textbf{Country} & \textbf{SEIR cases} & \textbf{DN-SEIR cases} & \textbf{True cases} & \textbf{$\epsilon_{r}$} & \textbf{$\epsilon_{rSEIR}$}\\
\hline
Belgium & 47,219 & 51,532 [45,130-60,464] & 47,859 &  0.07  & 0.01\\
France & 213,103 & 191,188 [166,043-224,554] & 128,442 & 0.33* & 0.66\\
Germany & 161,383 & 181,438 [163,333-207,151] & 157,641  & 0.13 & 0.02\\
Greece & 2,429 & 2,478 [2,343-2,666] & 2,576 & 0.03* & 0.06\\
Italy &6 201,802 & 202,369 [187,205-223,194] & 203,591 &<0.01* & <0.01\\
Luxembourg & 3,668 & 3,724 [3,578-3,936] & 3,769 & 0.01*\ & 0.03\\
Spain  & 209,646 & 230,794 [211,299-257,782] & 240,743 & 0.04* & 0.13\\
Brazil & 60,714 & 68,271 [55,268-86,164] & 78,162 & 0.14* & 0.22\\
Cameroon & 1,432 & 1,645 [1,375-2,003] & 1,832 & 0.11* & 0.22\\
Canada & 77,614 & 63,520 [51,977-78,913] & 51,597 & 0.19* & 0.5 \\
Japan & 12,250 & 11,353 [10,010-13,288] & 14,088 & 0.24 & 0.13\\
United Kingdom & 201,701 & 160,442 [134,251-188,088] & 165,221 & 0.03* & 0.22\\
\hline
\end{tabular}
\caption{Total cases as of 29/04 as predicted by a time-regression SEIR, and by DN-SEIR model. Both models trained until 11/04. $\epsilon_{r}$ and $\epsilon_{rSEIR}$  are the absolute relative error of the DN-SEIR model and time-regression SEIR respectively. (*) indicates that the DN-SEIR yields less error than the time-regression SEIR.}
\label{table:shortterm_fitting_cases}
\end{table*}



\begin{figure*}
\centering
\subfloat[Luxembourg]{\includegraphics[width=0.33\linewidth]{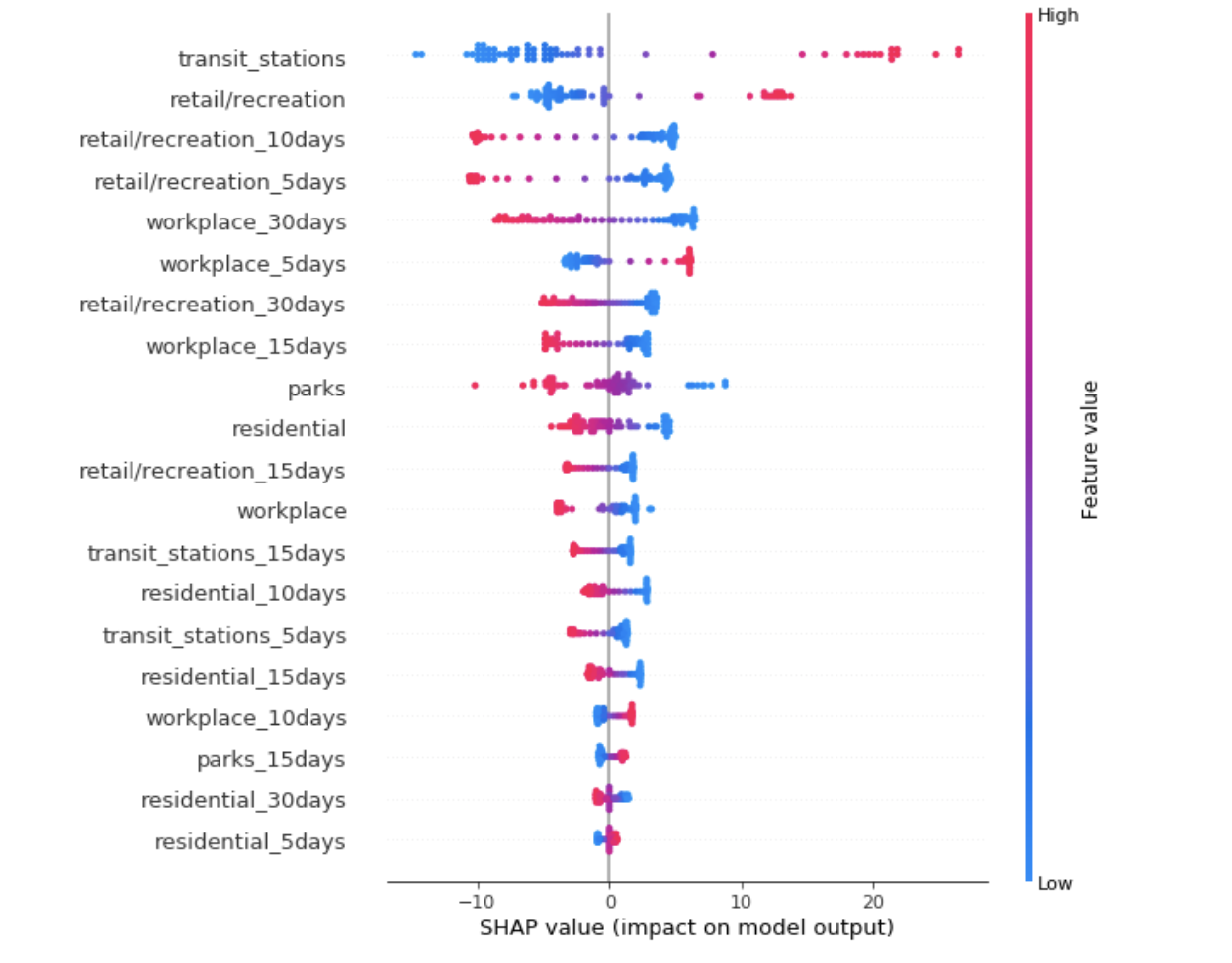} \label{fig:global_exp_luxembourg}}
\subfloat[Italy]{\includegraphics[width=0.33\linewidth]{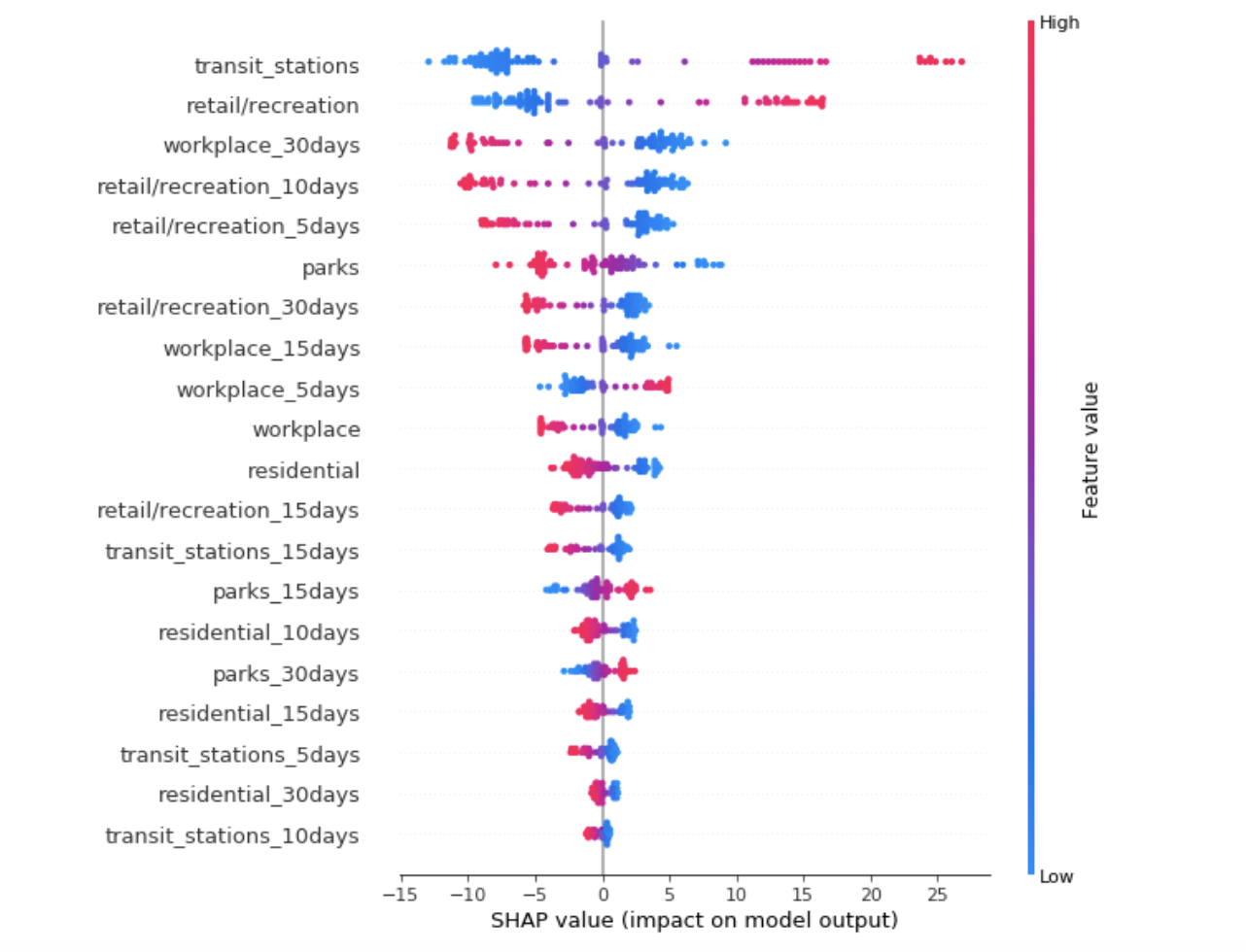} \label{fig:global_exp_italy}}
\subfloat[Japan]{\includegraphics[width=0.33\linewidth]{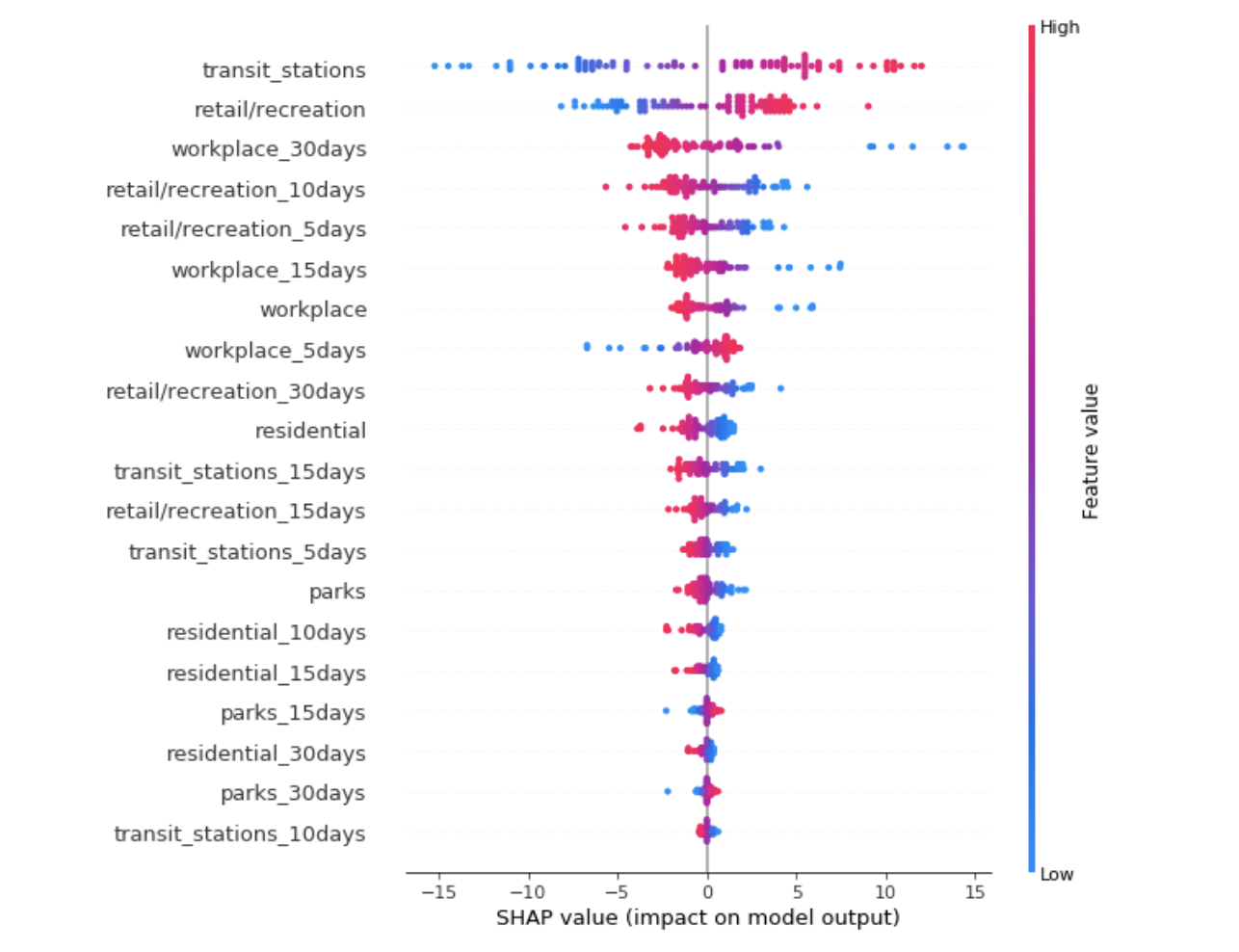} \label{fig:global_exp_japan}}
\caption{Global interpretation of the model for Luxembourg, Italy and Japan.}
\label{fig:interpretable_global}
\end{figure*}

\paragraph{Model interpretation at the global level} 
We fit \textit{SHAP} on the full dataset and show in Figure~\ref{fig:interpretable_global} the summary of its impact analysis where the features are ordered in decreasing order of influence. 
The distribution of each feature spans horizontally to inform on the impact of the feature on the final decision (positive on the right side), while the colour of each point of the distribution provides information on the range of values of the feature that have an impact (redder color codes for higher values of the feature). 
We can see for instance for the feature \textbf{transit station} that higher values have a positive impact on prediction, i.e higher values of \textbf{transit station} translate into higher values of \textbf{$R_t$}.  
This insight is common across the three countries. The same goes for the features related to \textbf{retail and recreation} activities.  

\textit{SHAP} shows how the trends in the feature, modelled in our approach using smoothed features, play a significant role in the final prediction, and shows an opposing contribution to its associated daily feature. For transit, retail, and park features, the trends values over 5, 10 and 15 days counter the impact of their respective daily values, yet on a lower scale. This indicates an inertia phenomenon that can be explained by the actual delay between the actual numbers ($R_t$, cases, deaths) and the reported one and also the delay inherent to the epidemiological model.

The comparison of the three countries also shows that mobility features have a much higher impact on the prediction in Italy and Luxembourg than Japan as their SHAP impact is much wider. This hints that other social distancing features in Japan could reduce the impact of mobility (masks for instance). 

\find{\textbf{RQ1 Answer:} Our approach yields predictions with much lower errors than pure epidemiological models in 75\% of the cases and achieves a 95\% R² score when the learning is transferred and tested on unseen countries. }

\subsection{Mid-term predictions with exit strategies}

In this section, we investigate four prediction strategies for an exit from the lockdowns that were taking place all over the world. The goal of the exit strategies is to allow a return to normal activities while minimizing the impact on the number of deaths and avoiding peaks in hospitalization that would saturate healthcare facilities. 
The strategies that we are investigating are the following:

\begin{itemize}
    \item \textbf{Hard exit}: In this strategy, all mobility activities are resumed to normal on May 11, 2020. 
    \item \textbf{Progressive exit}: Mobility activities are gradually restored, with an increase of 15\% of the activity every 2 weeks until the pre-lockdown activity level is reached.
    \item \textbf{Cyclic exit}: Every two weeks, activity is resumed to normal then brought back to lock down situation. The process is repeated for 4 cycles, thus ending on 03/08/2020.
    \item \textbf{Status Quo}: The current situation (as of April 30th) is maintained for the entirety of the period.
\end{itemize}

\begin{figure*}
\centering
\subfloat[Luxembourg]{\includegraphics[width=0.28\linewidth]{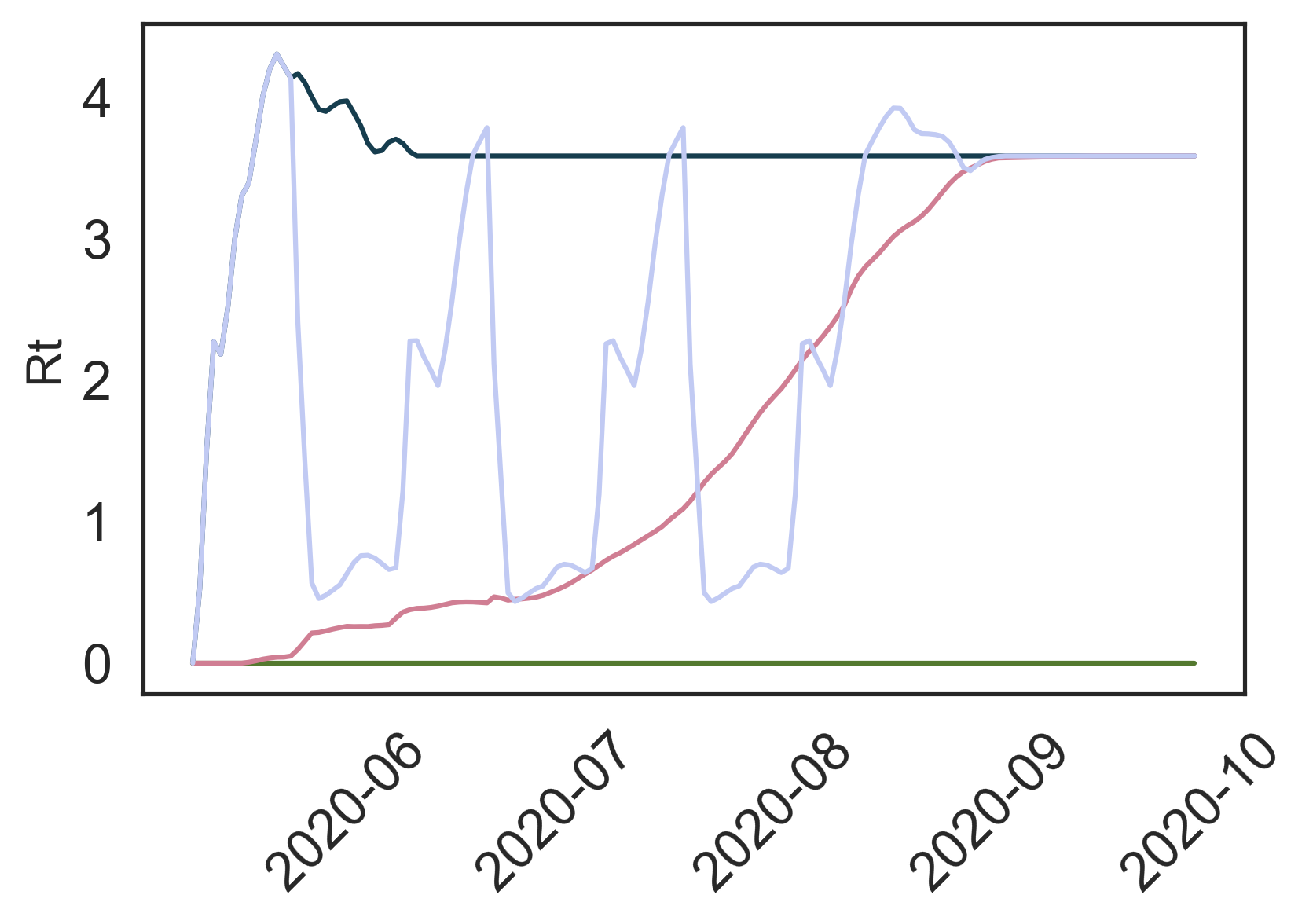} \label{fig:scenario_r_luxembourg}}
\subfloat[Italy]{\includegraphics[width=0.28\linewidth]{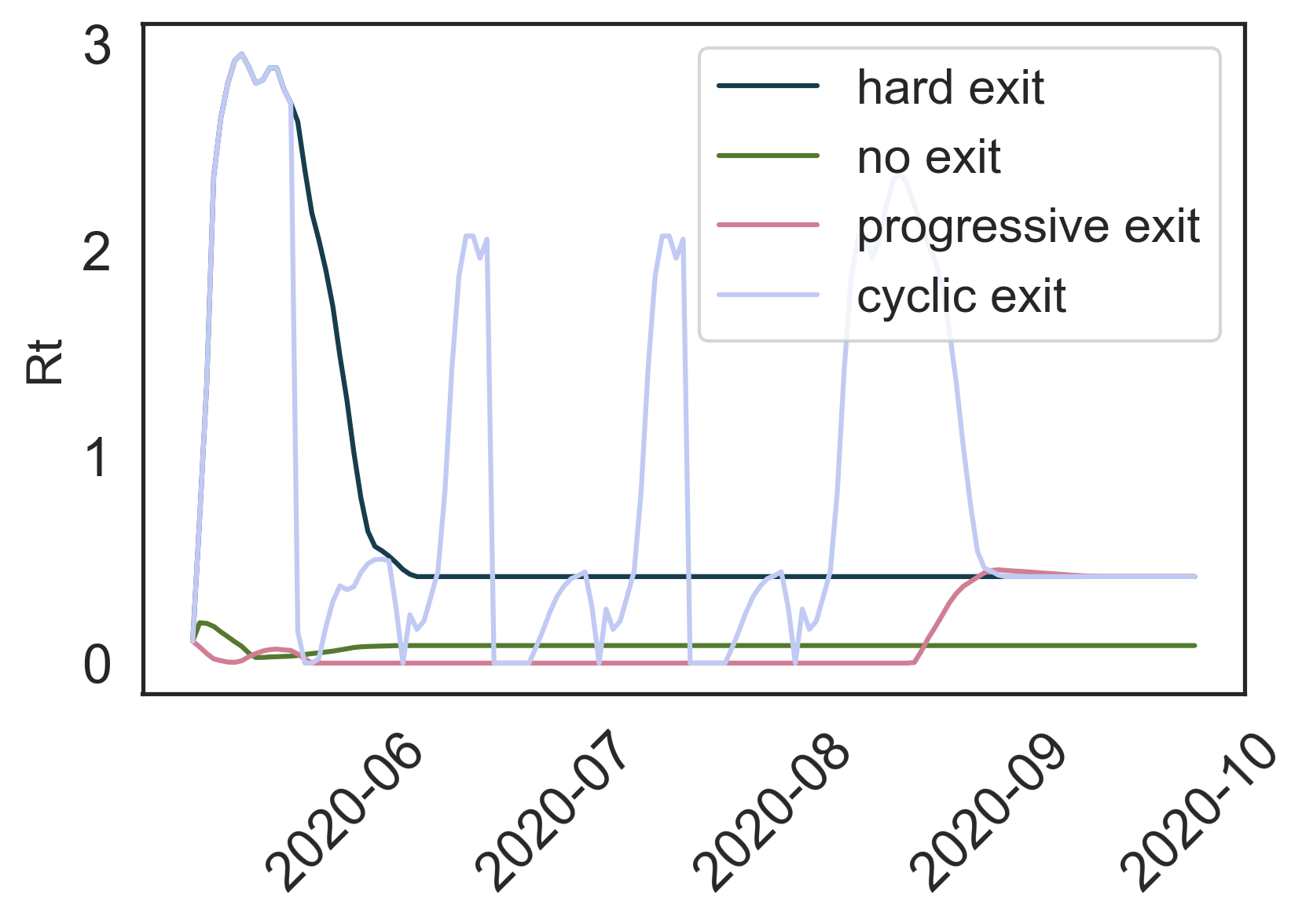} \label{fig:scenaria_r_italy}}
\subfloat[Japan]{\includegraphics[width=0.28\linewidth]{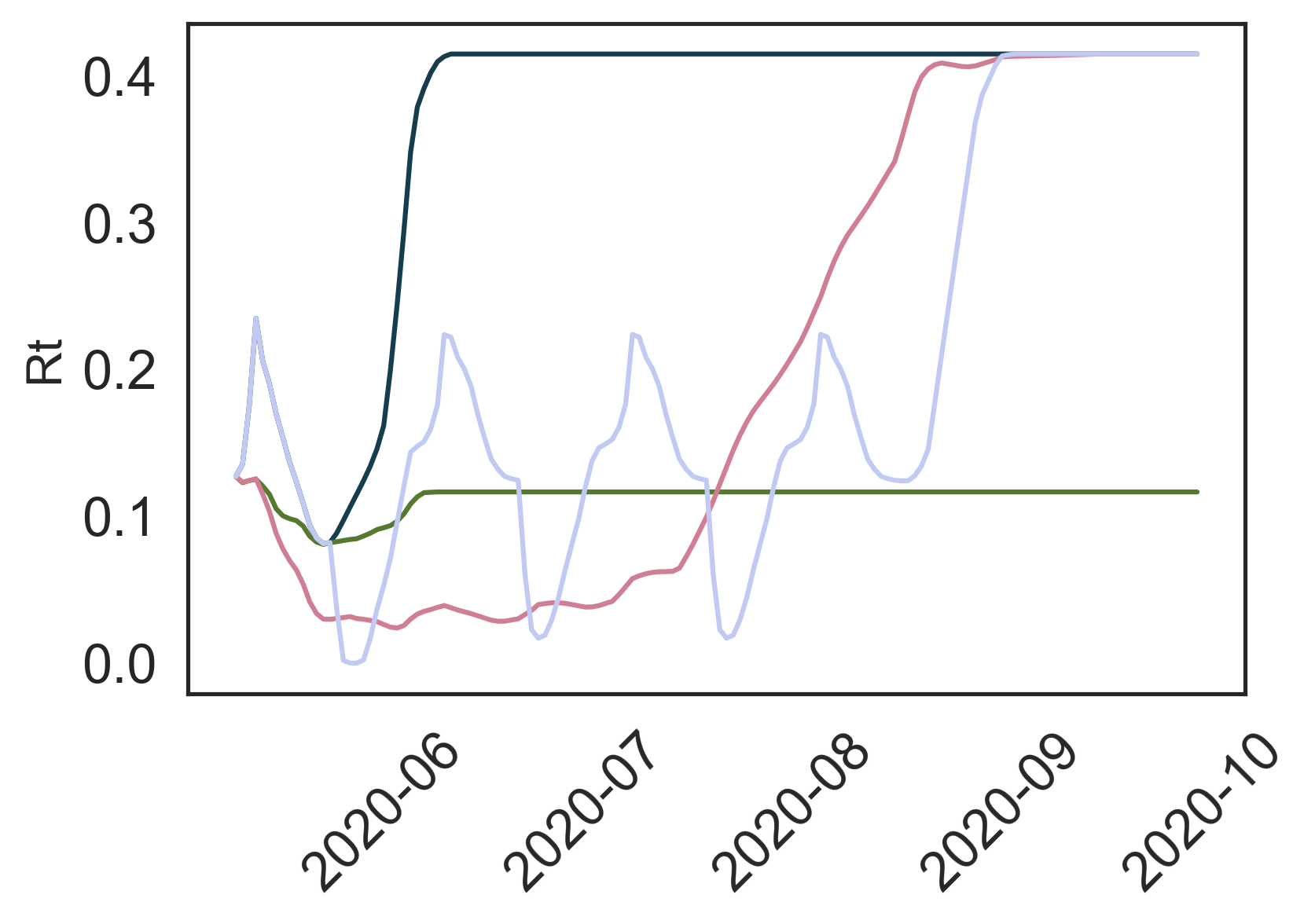} \label{fig:scenaria_r_japan}}
\caption{Evolution of the $R_t$ values for the four exit strategies modelled, \emph{i.e} a hard exit, a progressive exit, a cyclic exit and status quo for Luxembourg, Italy and Japan.}
\label{fig:scenario_r}
\end{figure*}

Figure~\ref{fig:scenario_r} shows the evolution of $R_t$ values for Luxembourg, Italy and Japan. As expected we see no evolution from the initial value in the no exit case, a cyclic fluctuation in the case of a cyclic exit, a soft increase of the $R_t$ values when applying a progressive exit and finally, a brutal jump in the case of a hard exit. $R_t$ reaches a plateau typically quite rapidly after strategies are applied. The plateau depends on the mobility condition, therefore, we see two plateaus in the results, one with all activities remaining closed (no exit), and one where all the other strategies reach the same plateau when the mobility levels are restored to their pre-lockdown baseline.

We evaluate these strategies for the three countries Luxembourg, Italy and Japan and obtain the results depicted by Table \ref{table:exit_strategies_mobility}. We choose those three countries because they present difference with respect to their demography, mitigation strategies and number of deaths attributed to COVID-19. Furthermore, they are amongst the few to provide reliable data about hospitalization capacities, hence allowing us to incorporate this information when looking for optimal policy schedules.

We compute for each strategy the Area Under Curve (AUC) of each mobility metric and provide the mean across all mobility values, and compare these hand-crafted strategies with the ones found by our Genetic Algorithm search. The search is run on 100 generations with a size of population of 100 and the hard constraint that critical hospitalizations should not exceed the country's ICU capacity (2,054 for Italy, 1,822 for Japan and 42 for Luxembourg, \cite{owidcoronavirus}). All strategies are evaluated between April 30th and September 30th.

We report 2 metrics, the total deaths on September 30th and the mean Area Under Curve across all the mobility features over the 5 months. The later reflects an economic objective that we need to maximize while the former is a healthcare objective. We state in the table three strategies found on the pareto. S1 is the pareto solution with the lowest death toll, S3 is the strategy with the highest mobility activity (and hence highest death toll) and S2 is the median death toll.

Our study shows that progressive lift strategies yield a similar economic footprint as 2-weeks cyclic strategies with fewer casualties (7\% fewer deaths for Italy and 10 times less for Luxembourg). 
\find{\textbf{RQ2 Answer:} Our approach allows to see drastic changes based on different exit strategies. The progressive strategy offers in our experiments a better outcome than a hard or a cyclic strategy.}

The search-based strategies (S1, S2, S3) perform better than manual strategies on the death metric for Italy and Japan, and for Luxembourg. For Luxembourg, S1 performs as good as the progressive strategy. 

Overall, our results show that the search for exit strategies can be guided and restricted to the policy-makers constraints (i.e. hospital capacity) and yields actionable strategies within constrained computation time. We ran the experiments 100 times (x100 generations each) and the hypervolume of the pareto-solutions converges within 80 generations. 

\find{\textbf{RQ3 Answer:} The search algorithm yields better strategies than the popular ones both in term of impact on the activity and number of deaths while ensuring that the healthcare facilities are not overwhelmed.}

\begin{table}[h]
\resizebox{0.8\linewidth}{!}{
\begin{tabular}{|l|l|r|r|}
\hline
\textbf{Country} & \textbf{Strategy} & \textbf{Mobility AUC} & \textbf{Deaths}\\
\hline
Luxembourg & Status Quo     & -10,721.6 & 108 \\
           & Hard           & -91.9     & 2,763 \\
           & Progressive    & -2,774.43 & 114 \\
           & Cyclic         & -2,487.02 & 2,002 \\
           & Pareto-S1      & -2,381.45 & 165 \\
           & Pareto-S2      & -2,370.7  & 635 \\
           & Pareto-S3      & -2,289.7  & 697 \\
\hline
Italy      & Status Quo     & -6,006.13 & 32,015 \\
           & Hard           & -57.53    & 37,377 \\
           & Progressive    & -4,124.93 & 31,987 \\
           & Cyclic         & -3,689.13 & 34,427 \\
           & Pareto-S1      & -8,412.125& 29,449 \\
           & Pareto-S2      & -7,570.0  & 29,450 \\
           & Pareto-S3      & -7,275.38 & 29,452 \\
\hline
Japan      & Status Quo     & -3,431.7  & 709 \\
           & Hard           & 14.52     & 710 \\
           & Progressive    & - 1,005.3 & 708 \\
           & Cyclic         & -896.03   & 709 \\
           & Pareto-S1      & -2,106.21 & 654 \\
           & Pareto-S2      & -1,170.33 & 660 \\
           & Pareto-S3      & -1,106.33 & 671 \\
\hline

\end{tabular}
}
\caption{Exit strategies comparison. Higher AUC and lower deaths are better.}

\label{table:exit_strategies_mobility}
\end{table}

\section{Conclusion}

In this paper, we studied \emph{DN-SEIR}, a data-driven approach to evaluate the effective reproduction number of the COVID-19 epidemic. In particular, we considered both manual and search-based mitigation strategies, with the aim to help decision-makers in the evaluation and selection of exit strategies. To this end, we evaluated the state-of-the-art compartment model (i.e. SEIR) and shew that our approach yields predictions closer to the ground truth. We also demonstrated that learning can transfer across different countries and a simple FFNN provides accurate and interpretable predictions. Finally, we proposed a search-based approach to evaluate and find optimal strategies that satisfy the constraints of the health facilities and achieve a quick economic recovery with limited casualties.

Our approach paves the ways to automated strategy simulation and search and provides a simple, yet, powerful tool for policy makers to tailor exit strategies to their context and priorities. We can go further than our approach with better feature engineering or neural architecture search (with CNN or RNN). We can also extend the data-driven prediction of hyper-parameters not only to the effective reproduction number but also to all the epidemiological parameters like hospitalization rate. This would require having access to accurate hospitalization data across a large pool of countries and can be achieved in the close future as more countries are sharing such data. Finally, we could extend our technique to a more-grained approach that takes into account age-specific or location-specific epidemiological models.

\bibliographystyle{ACM-Reference-Format}
\bibliography{references}

\end{document}